\documentstyle[twoside,fleqn,epsfig,espcrc2]{article}

\def\Journal#1#2#3#4{{#1} #2 (#3) #4}

\def\PLB{{Phys. Lett.}  B}
\def\PRL{Phys. Rev. Lett.}
\def\PRD{{Phys. Rev.} D}

\def\APH{Astropart.Phys.}

\def\be{\begin{equation}}
\def\ee{\end{equation}}
\def\bea{\begin{eqnarray}}
\def\eea{\end{eqnarray}}
\newcommand{\AmS}{{\protect\the\textfont2
  A\kern-.1667em\lower.5ex\hbox{M}\kern-.125emS}}

\title{Neutrino telescopes under the ocean: The case for ANTARES}

\author{L. Moscoso\address{DSM/DAPNIA/SPP, CEA/Saclay, \\
               91191 Gif-Sur-Yvette CEDEX, France}%
        \thanks{On behalf of the ANTARES collaboration}
        }
       
\begin{document}

\begin{abstract}
Neutrino telescopes offer an alternative way to explore the Universe.
Several projects are in operation or under construction. 
A detector under the ocean
is very promising because of the very accurate angular resolution
that it provides. The ANTARES project is intended to demonstrate
the feasibilty of such a detector.
\end{abstract}

\maketitle

\section{INTRODUCTION}

High energy cosmic neutrinos should provide a new means to explore the sky. 
Due to the weakness of their interaction with matter and the absence of 
their 
interaction with the electromagnetic radiations, neutrinos can travel in the 
Universe without being absorbed. On the contrary, because photons interact 
with matter, with the infra-red (IR) radiation and with the cosmic microwave 
background (CMB), the Universe is opaque for high energy gamma rays.

Charged cosmic rays such as protons are deflected by the galactic and 
extra-galactic magnetic fields. So, only UHE protons, above $10^{20}$\,eV, 
are rigid enough to point back to their source. Nevertheless, at these 
extreme energies protons also suffer from interactions with IR radiations 
and with the CMB which limit their free path-length to about 50\,Mpc. 
At greater distances protons continuously lose energy and become less and 
less rigid.

Therefore, it appears that the only way to explore the Universe in the 
high energy range and at great distances is to detect neutrinos.

\section{DETECTION OF HIGH ENERGY NEUTRINOS}

High energy muon neutrinos can be detected by searching for long-range muons 
produced in charged current exchange
interactions of neutrinos with the matter
surrounding the detector. Due to the increase of the $\nu N$ cross-section 
with the neutrino energy and to the increase of the muon path-length with 
the muon energy, the probability to detect a muonic 
neutrino aimed towards the 
detector is an increasing function of the neutrino energy. This means that 
high energy neutrinos will be statistically enhanced. Moreover, 
the angle between the neutrino and the produced muon is very small for high 
energy neutrinos. So the direction of the parent neutrino is well
determined. 

Despite its increase with the neutrino energy, the $\nu N$ cross-section 
remains small and, moreover, the neutrino flux is expected to be
a decreasing function of the neutrino energy with a differential
$\simeq$E$^{-2}$ behaviour.
The detector area must be large enough to provide sufficient  
sensitivity to detect cosmic sources over the widest possible solid angle
on a reasonable time scale. 
For this reason a volume of detection of 1\,km$^3$ is needed.

At the surface of the Earth the main source of background is the flux of 
particles
produced in the cascades initiated in the atmosphere by primary cosmic
rays. The major component is downward-going muons which can be rejected
by selecting only upward-going particles. In order to suppress
particles produced in the back-scattering of atmospheric muons, the detector 
must be well shielded. The remaining source of physical background is
the flux of upward-going neutrinos produced in the atmospheric cascades.

The most economic way to realise a km-scale well-shielded detector is to
 build a 3-D array of optical modules in the deep ocean or in polar ice. 
In these media high energy muons crossing the detector produce 
\u{C}erenkov light at an angle ${\theta}_C \simeq 43^{\circ}$. The 
reconstruction of the muon direction is performed by using the information 
on the arrival times of the photons recorded by the optical modules. 

\section{THE ANTARES PROJECT}

The main advantages of deploying the detector in the ocean compared to 
ice are the long scattering length and low scattering angle of light in
water, the possibility to deploy on different sites located at different
latitudes and the possibility to find sites at great depth. Nevertheless,
several questions must be answered. In particular tests are needed to
master the deployment and connection operations in the 
deep sea. Moreover the environmental parameters like the optical
background rate, the bio-fouling rate and the water transparency must
be measured.
The ANTARES project~\cite{antares} has two goals:
\begin{enumerate}
\item Realisation of apparatuses capable of measuring environmental 
parameters such as optical background, bio-fouling and water transparency;
\item Construction and deployment of a 3-D prototype (``Demonstrator'') 
scalable to a cubic kilometer detector.
\end{enumerate}
These operations are performed off Toulon (France), 30\,km from the 
shore at 2350\,m depth. When the feasibility of a large detector has been 
demonstrated, further steps towards a cubic kilometer will be proposed.

\subsection{The demonstrator}
 
The deployment of a large network of optical modules in the deep ocean is 
one of the major problems to be solved. A very detailed programme has been 
defined in the ANTARES project to proceed by steps in order to ensure the 
good quality of the procedures and to reduce causes of failures.

The final prototype will consist of three strings, equipped with about 30 
optical modules each, electrically interconnected through a junction box 
which will be linked to the shore via an electro-optical cable. The signals 
delivered by each optical module will be transmitted to the shore station 
through the optical fibers of the electro-optical cable.

The 40\,km long electro-optical cable was successfully deployed in May 98.
Mechanical tests of the first string started in July 98 and 
finished in September. Connection-deconnection tests are foreseen for the
end of 98. A first string equipped with 8 optical modules, electronics,
and positioning and slow-control systems will be deployed by the
end of 98 and connected to the electro-optical cable. The first fully
equipped string is foreseen for 99. 

\subsection{Environmental parameters}

The site where the final very large detector will be installed must have 
very good properties from all points of view. The ANTARES collaboration is 
constructing a system of autonomous detectors capable of measuring the sea 
quality at any depth down to 4\,000\,m in order to chose the site and the 
characteristics of the final detector.

\subsubsection{Optical background}

The optical background has two main components: 
a continuous component due to 
\u{C}erenkov light produced by electrons emitted in the $\beta$ decay 
of the $^{40}$K present in the sea water and a variable component due 
to the bio-luminescence emitted by bacteria and fishes.

\begin{figure}[htb]
  \mbox{
     \epsfig{file=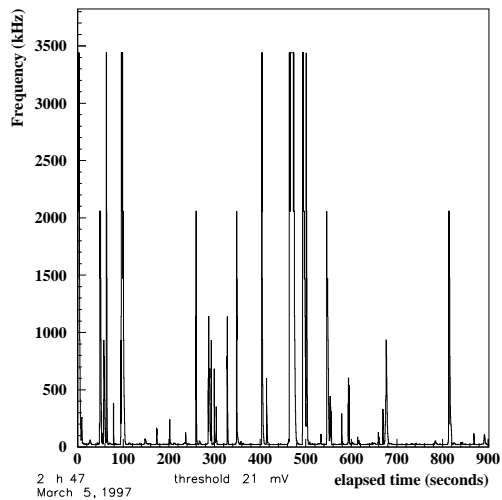,
      width=\linewidth}}
\caption{Variations of the counting rate with time.}
\label{fg:biolum}
\end{figure}

\begin{figure}[htb]
  \mbox{
     \epsfig{file=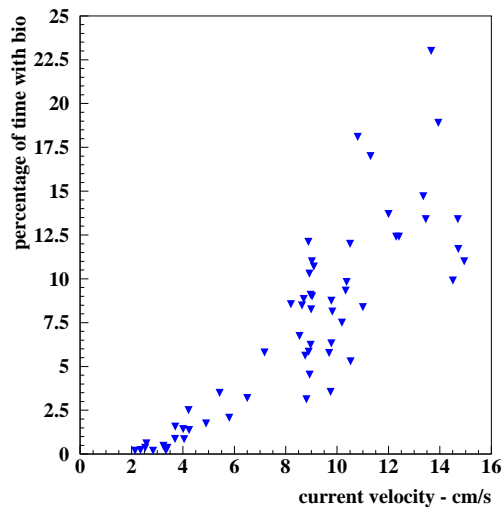,
      width=\linewidth}}
\caption{Percentage of time where the counting rate exceeded by 15\% the 
$^{40}$K background vs the current speed.}
\label{fg:biovscurr}
\end{figure}

Several deployments of 350\,m long strings at 2\,350\,m depth have been 
performed. Each string supports several PMTs (1-3) and other monitors such as 
currentmeters, thermometers, compasses and tiltmeters. Each string is 
equipped with a data-logger, an acoustical modem to transmit data to 
the boat, and a system of releases for recovery. The power is supplied 
by lithium batteries. Figure~\ref{fg:biolum} shows an example of the
variations of the counting rate recorded with our system during
a period of medium activity.
The pulse height threshold used to measure the counting rate was set to a
value corresponding to 1/3 of the mean amplitude for a single photo-electron.
The figure clearly shows the bio-luminescent activity consisting of bursts
of light lasting a few seconds. The continuous background due to the $^{40}$K
decay is also visible with a frequency of $\approx$~40\,kHz. The 
fraction of time during which the counting rate exceeded the $^{40}$K 
background by at least 15\% was found to be strongly correlated to the current 
speed, as shown in figure~\ref{fg:biovscurr}.

The measurement performed simultaneously with two PMTs 40\,m apart
showed a very little 
time correlation, suggesting that the size of the region where the
bio-luminescence is active is generally less than 40\,m.

\subsubsection{Bio-fouling}

The measurement of the bio-fouling of the glass sphere which will house the 
PMT is a long term operation. It has been performed twice for periods of
3 and 8 months. The system was a string equipped with two glass spheres; 
one of them (A) housed  
a system of light diodes (LED) which were continuously monitored, and the other
one (B) housed several PIN diodes located at different lattitudes of the 
sphere.
During the first deployment the frame supporting the spheres 
was oriented vertically with the sphere A
in the upper position emitting light downwards towards the sphere B. 

\begin{figure}[htb]
  \mbox{
     \epsfig{file=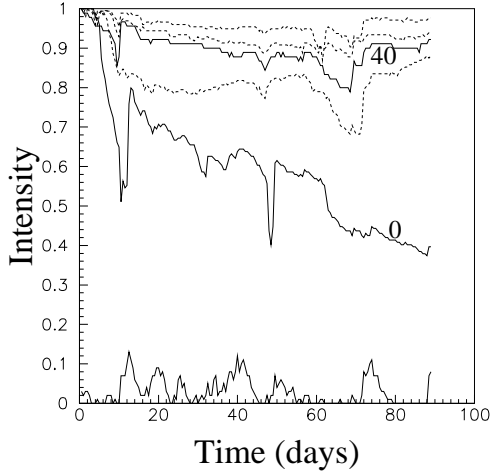,
      width=\linewidth}}
\caption{Evolution of the PIN diode counting rates measured
during the first period of 3 months with the frame supporting the spheres
in the vertical position. The upper
curves depict the counting rates of PIN diodes at 
different lattitudes. Full lines are for PIN diodes at zenith angles 
of 0$^{\circ}$ (top of the sphere)
and 40$^{\circ}$ as indicated by the labels. Dashed lines are for PIN diodes
at 20$^{\circ}$. The intensities are normalised to the first measurement. 
The curve on the bottom of the figure shows the
variation of the water current velocity in m/s.}
\label{fg:fouling1}
\end{figure}

The result of this measurement, of 3 months duration, 
is shown in figure~\ref{fg:fouling1}.

The counting rate of upward-facing PIN decreased by 60\% in 3 months while 
the rates of the PINs
located at other lattitudes decreased less. The counting rates
are highly discontinuous and increase suddenly in the presence of high
water current velocity. This suggests that {\em i)} the bio-fouling covered
only a small angular region on the top of the glass sphere and {\em ii)} the
bio-fouling was partially removed by the water current probably because
it was not strongly glued on the glass.

\begin{figure}[htb]
  \mbox{
     \epsfig{file=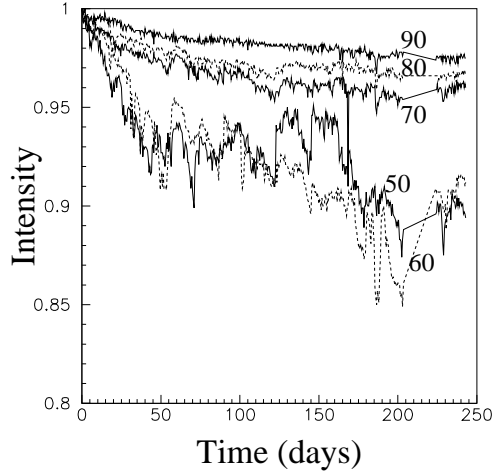,
      width=\linewidth}}
\caption{Evolution of the PIN diode counting rates measured
during the second period of 8 months with the frame supporting the spheres
in the horizontal position. Different
curves depict the counting rates of PIN diodes at 
different lattitudes. The zenith angle (degrees) of the position of the 
PIN diode 
is indicated by the label on each curve. 
The intensities are normalised to the first measurement. Note that the 
vertical scale is different from that of figure~\ref{fg:fouling1}. }
\label{fg:fouling2}
\end{figure}

A second measurement was performed after orienting the frame horizontally.
The result of this 8-month-long 
measurement is
shown in figure~\ref{fg:fouling2}. Taking into account that the LED sphere 
suffers the same fouling as the PIN sphere, one can estimate that the 
horizontal region of the glass sphere loses only about 1.5\% of
transparency in one year.

\subsubsection{Water transparency}

The light attenuation as a function of the distance from the source was
measured using
a 350\,m long mooring string incorporating a 33\,m long rigid structure 
holding an optical module at one end facing a motorised trolley carrying a 
light source along the structure. Continuous 
light sources (LEDs) emitting at different wavelengths are used. The 
measured attenuation length is 39$\pm$3\,m for a wavelength
of 466\,nm.
In July 98 
a pulsed light source was used to disentangle the contributions
from light absorption and from light scattering. The analysis is in
progress. 

\subsection{Optical modules}

Different large photocathode photomultipliers are being tested (EMI and 
Hamamatsu 8" tubes). Larger tubes (10" and 11") are also 
foreseen. A dark box equipped with a mechanical system allowing a blue LED 
to scan the photocathode area is used to measure the response of the 
phototubes as a function of the position of the light spot. A water 
tank is used to study the overall response of the optical module to 
the \u{C}erenkov light emitted by cosmic muons in water.

\subsection{Data transmission}

The optical module signals will be transmitted to the shore through the 
optical fibers of the electro-optical cable. Analog data transmission will 
be used first and replaced later by digital transmission of data from a
signal sampling device based on an ASIC 
chip currently under development. The 
electro-optical cable, deployed in May 98, is
equipped with four mono-mode fibers. The measured attenuation is 
0.33\,dB/km at a wavelength of 1\,310\,nm.

\subsection{Positioning and slow control}

The knowledge of the relative position of the optical modules should match 
the size of the photocathodes of the PMTs (20\,cm). This will be 
achieved by sonar triangulation between an external base and acoustic 
detectors along the strings. Adjustment of the PMT voltages, 
recording of the environmental parameters, and measurement of 
the detector geometry will be managed by the 
slow-control system. These monitoring data will be transmitted to the 
shore through the electro-optical cable.

\subsection{Software}

A software package has been developed to simulate the neutrino interaction in 
the medium surrounding the detector, the muon tracking, the \u{C}erenkov 
light emission and the detector response. The optical background and the 
distortions of the detector by the water currents are also simulated. 

\section{TOWARDS A CUBIC KILOMETER DETECTOR}

The development of a cubic kilometer detector is a very complex challenge
which must be reached by steps. The ANTARES collaboration envisages an
intermediate stage consisting of a detector
of 0.1\,km$^2$ made of a
network of about 1000 PMTs. The software packages developped for the 
ANTARES project have been used to estimate the performance expected for
this intermediate stage. Different detector layouts have been considered, all
made of 15 strings equipped with a total of about 1000 PMTs. The results are
quite similar for the different layouts.

\begin{figure}[htb]
  \mbox{
     \epsfig{file=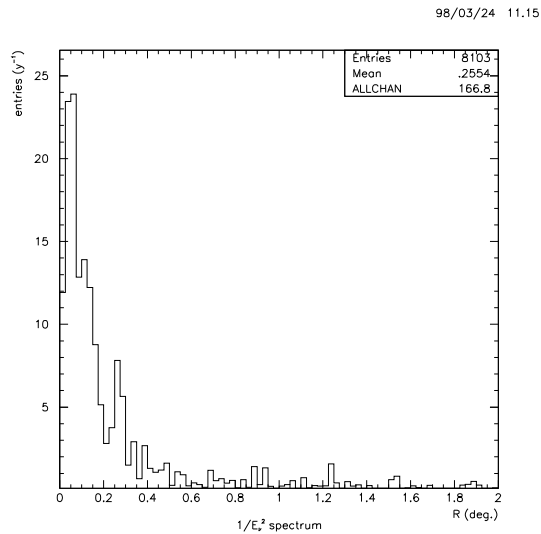,
      width=\linewidth}}
\caption{Distribution of 
the angle between the direction of the muon entering the detector and the
direction of the reconstructed track for muons induced by simulated neutrinos
with a E$^{-2}$ differential energy spectrum.}
\label{fg:resol}
\end{figure}

The 
reconstruction algorithm applied to simulated events shows that the angular 
resolution for reconstructed tracks will be better than 0.2$^{\circ}$. 
Moreover, for muons with energies that trigger a km-scale detector 
(above 1-10\,TeV), the angle between the muon and the parent neutrino is 
smaller than 0.1$^{\circ}$. Figure~\ref{fg:resol} shows the distribution of 
the angle betwen the direction of the muon entering the detector and the
direction of the reconstructed track for muons induced by simulated neutrinos
with a E$^{-2}$ energy spectrum.

\section{EXTRAPOLATION OF THE EGRET CATALOGUE}

Powerful extra-galactic objects such as 
AGNs or topological defects could contribute to a diffuse flux of
high energy neutrinos 
significantly larger than the flux of atmospheric 
neutrinos at high energy~\cite{agn,td}.

Moreover, due to the extremely low angle between the muon and the parent 
neutrino and to the good quality of the muon direction measurement, the 
atmospheric neutrino background contaminating each individual source can be 
reduced to a very low level by selecting very small angular regions of the 
sky. In that way, a signal of only a few events could be significant.
The sensitivity of the detector to muon neutrinos can be estimated from 
measured low-energy gamma-ray fluxes by assuming that {\it i)}
 the low energy gamma-rays are of hadronic origin and {\it ii)} the emitted 
gamma-rays have a differential energy spectrum $E^{-2}$.

With these assumptions the muon neutrino flux is about 40\% of the flux of 
gammas at the production source. Using the 2nd EGRET catalog~\cite{egret:cat} 
for sources measured during the P12 period, the derived neutrino flux has been 
extrapolated to the energies where the neutrino detector is sensitive. 

Although no individual extra-galactic source
can be detected with an exposure of 0.1\,km$^2 \cdot$year, a statistically
significant effect can be detected by adding the contributions of all the
extra-galactic sources. A calculation made for the 43 identified AGN gives
a total number of 8-67 events (depending on the value of the differential
spectral index used: respectively 2.2 and 2) to be compared to a total
background of 2.7 events.

A possible scenario would be to start with a 0.1\,km$^2$ detector 
made of about 15 strings equipped with about 1000 optical modules. If a
statistical enhancement correlated
with the positions of the AGN sources could be detected on a time scale of 
the order of one year this would motivate 
the construction of the cubic kilometer detector.

\section{CONCLUSIONS}

The sky survey with high energy neutrinos is essential in order to
obtain new information complementary to that obtained from low energy 
gamma-rays and short
distance high energy gamma-rays and ultra high energy protons. This requires a 
detector at the kilometer scale which can only be developed in stages.
The construction of a deep ocean prototype and a programme of measurement
of environmental parameters is the first step necessary to
demonstrate the feasibility of such a detector.

Studies performed by the ANTARES collaboration indicate that a detector in the
ocean can determine the origin of each event with a very fine angular
resolution. This is a major advantage of the underwater technique, making
it possible to reduce the background from known point sources to very low
level.

\end{document}